\documentclass[sigconf,10pt]{acmart}

\usepackage{booktabs} 

\setcopyright{rightsretained}

\usepackage{graphicx}
\usepackage{listings}
\usepackage{subcaption}

\acmDOI{10.475/123_4}

\acmISBN{123-4567-24-567/08/06}

\acmConference[UCF]{}{April 2019}{Orlando, Florida} 
\acmYear{2019}
\copyrightyear{2019}

\acmPrice{15.00}

\begin{document}
\title[Analysis of Commutativity]{Analysis of Commutativity with State-Chart Graph Representation of Concurrent Programs}

\author{Kishore Debnath}
\affiliation{%
  \institution{University of Central Florida}
  \streetaddress{4000 Central Florida Blvd.}
  \city{Orlando} 
  \state{Florida} 
  \postcode{32816}
}
\email{kishore.debnath@knights.ucf.edu}

\author{Christina Peterson}
\affiliation{%
  \institution{University of Central Florida}
  \streetaddress{4000 Central Florida Blvd.}
  \city{Orlando} 
  \state{Florida} 
  \postcode{32816}
}
\email{clp8199@knights.ucf.edu}

\author{Damian Dechev}
\affiliation{%
  \institution{University of Central Florida}
  \streetaddress{4000 Central Florida Blvd.}
  \city{Orlando}
  \state{Florida} 
  \postcode{32816}
}
\email{dechev@cs.ucf.edu}

\renewcommand{\shortauthors}{K. Debnath et al.}

\begin{abstract}
We present a new approach to check for commutativity in concurrent programs from their state-chart graphs. A set of operations are commutative if changing the order of their execution on an object does not affect the abstract state of the object and returns the same response. Concurrent operations that commute at object-level can be executed concurrently at transaction-level, which boosts performance while preserving the appearance of atomicity and isolation. Utilizing object-level commutativity in transactional execution enables the reuse of existing non-blocking programming techniques for thread-level synchronization. In our approach, we generate state-chart graphs by tracking data on the atomic instructions invoked on the concurrent object during model checking and represent the atomic instructions as states in a state-transition representation. Considering the non-deterministic nature of concurrent programs, we determine commutativity by exhaustively searching for identical object states captured at a thread-level granularity across all thread interleavings. With this methodology, a user can not only verify commutativity among operations, but also can visually check ways in which methods commute at object-level, which is an edge over current state-of-the-art tools. The object-level commutative information helps in identifying faulty implementations and performance improvement considerations. We use the graph database, Neo4j, to represent object states as nodes that further assists the user to check for concurrency properties using Cypher queries. 
\end{abstract}

%
%

\ccsdesc[500]{Computing methodologies~Concurrent programming languages}
\ccsdesc[500]{Computing methodologies~Concurrent algorithms}

\keywords{Commutativity, State-chart, Graph, Graph database, Concurrent programs}

\maketitle

\section{Introduction}

In the past decade, there has been a paradigm shift in high-performance programming as Moore's law began to diminish for single core processors. The quest for high performance scalable solutions has compelled developers to make better use of the new class of processors comprising multiple cores. As developers started to focus on multithreaded programming models, non-blocking programming soon became a popular design choice over poorly scalable blocking solutions. Software Transactional Memory (STM)\cite{herlihy2008transactional}\cite{herlihy1993transactional}, an alternative to traditional mutual exclusion constructs, emerged as a scalable solution for concurrent objects. Herlihy proposed Transactional Boosting \cite{herlihy2008transactional} as a technique for transforming concurrent linearizable objects into concurrent transactional objects. From the perspective of concurrent programming, a transaction can be considered as a composition of one or more concurrent operations. 
Two operations commute if applying them in either order leaves the object in the same state and returns the same response \cite{herlihy2008transactional}.
A higher degree of concurrency can be achieved by allowing commutative operations to execute simultaneously in transactions. 
When commutative operations in two separate transactions are allowed to proceed concurrently through thread-level synchronization, performance is improved due to the elimination of unnecessary transaction-level synchronization for low-level memory accesses. 

\begin{figure*}[t]
\centering
\includegraphics[width=\textwidth]{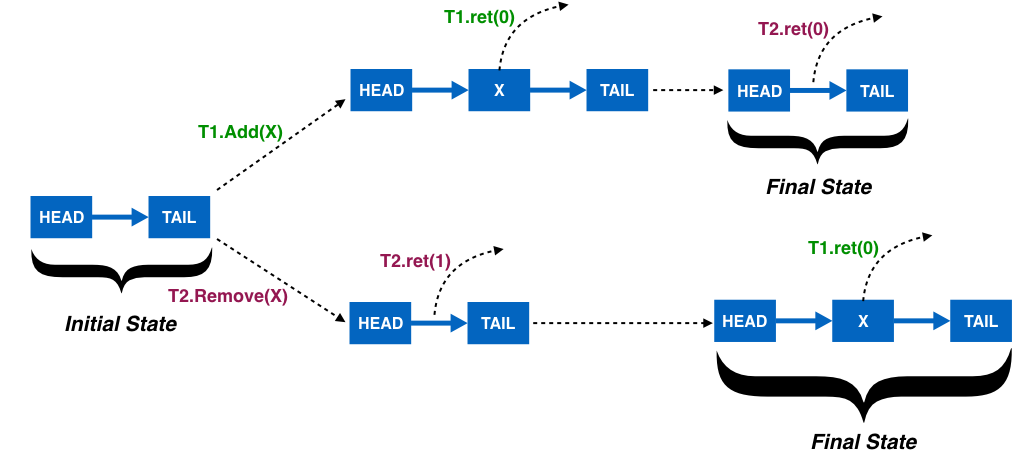}
\caption{Linked list object that is initially empty gets concurrently executed by add($X$) and remove($X$) operation. In one trace, the add($X$) operation precedes the remove($X$), and in the final state the linked list remain empty. Conversely, another trace where remove($X$) preceded add($X$) leaves the linked list in a different state.}
\label{ll_1_add_1_remove}
\end{figure*} 




The design and implementation of correct concurrent programs is a challenging task. The verification of proven scalable properties like commutativity during implementation phase is essential to ensure that the concurrent program is operating according to its design. Previous work by Clements et al. \cite{clements2015scalable} checks commutativity in interface operations, where two operations are considered to commute if they are \textit{conflict-free}; that is, no core writes to a cache line that was read or written by another core. This definition of commutativity does not account for object-level operations that commute even in the presence of memory access conflicts.


In this paper, we present a technique that uses a state-transition graph to check commutativity between object-level operations. For this purpose, we designed Concordant, a tool that runs on top of any model checker and converts model checking data of a concurrent program into states and transitions. Model checking data such as thread create, thread finish, atomic read, atomic write, method invocation, method response, etc. are represented as action nodes in state-chart graphs. Here we use the terms ``action nodes" or ``states", and ``transitions" or ``relationships" interchangeably. Graphical state-transition representation of a concurrent program provides comprehensive information of all concurrent operations, which is missing in any state-of-the-art tools solving similar problems. 

With our state-chart notation, we capture instruction details such as the type of instruction (in our case mostly atomic operations) in execution, input/response value, memory location in execution, etc., which suffices with required information used for debugging a multithreaded program. Action node ordering in our state-chart conveys information such as how atomic operations interleave across multiple execution traces. We \textit{probe} memory locations by calling object operations after the concurrent execution in order to evaluate the final object state in the state-chart. By probing memory locations of object operations and representing object states into nodes, we help developers compare program behavior across multiple traces and investigate error in the code.  

A significant amount of effort and ingenuity has been dedicated to the development of non-blocking synchronization techniques. Knowledge of commutative operations allows the reuse of these techniques for thread-level synchronization in transactional execution. Additionally, a significant performance increase can be achieved by allowing commutative operations in transactions to proceed concurrently in a thread-safe manner. Concordant enables the user to identify commutative operations and exploit object semantics when designing programs that require the execution of a composition of operations in the form of a transaction.

We present in Figure~\ref{ll_1_add_1_remove} a sample of non-commutative operations in the set abstract data type implemented as a concurrent linked list from \cite{herlihy2011art}. Although the remove($X$) operation commutes with any ordering of an add($Y$) operation given that $X \ne Y$, the operations are no longer commutative when $X = Y$. 
Consider a linked list is initially empty and an add($X$) operation trying to insert \textit{X} and a remove($X$) operation is competing to remove \textit{X} concurrently during execution. If the add($X$) operation precedes the remove($X$) operation in an execution, then the linked list remains empty in the final state. Whereas, if the remove($X$) precedes the add($X$) operation then the linked list will contain the element \textit{X} at the end of the execution, shown in Figure~\ref{ll_1_add_1_remove}. Evidently, these two operations cannot commute as the final states do not remain the same at object-level across all execution traces. We verify this non-commutativity between add($X$) and remove($X$) by evaluating atomic action nodes in our state-chart graph. The deviations in the final state represented by action nodes help us to deduce that add($X$) and remove($Y$) do not commute when $X = Y$. 

Our idea of state-chart diagram is derived from the notion of trace theory concepts like dependency-graphs and concurrent histories \cite{mazurkiewicz1995introduction}\cite{hoogeboom1995dependence}, which forms the mathematical model to evaluate properties of concurrent systems. Using trace theory, we explore different commutative instances using abstract symbols, which we then apply in our state-charts to derive the concept of \textit{preconditional} and \textit{postconditional} states. A preconditional state is a state of an object prior to the concurrent execution of operations being evaluated for commutativity. A postconditional state is a state of an object after the concurrent execution of operations being evaluated for commutativity. Preconditional and postconditional states are verification points to check commutativity in methods. We use the graph database Neo4j \cite{Neo4j} to represent concurrent model states as nodes and model checker properties as node properties. This technique for commutativity checking has not been done in any previous state-of-the art tools. The advantage of using a graph database is that we can further analyze each of the commutative methods across all traces using the Cypher \cite{Cypher} query language. State-chart graph representation of concurrent program opens a new research discourse to analyze different characteristics of non-blocking programs, some of which are presented in this paper.  

This paper makes the following contributions:

\begin{itemize}

  \item We present Concordant, a tool for analyzing commutative operations by generating a state-chart graph comprising all traces explored during model checking. Atomic operations are represented as action nodes in the state-chart graph.
  
  \item We present the application of Concordant to verify commutativity in two data structure classes: contiguous and non-contiguous memory. We achieve this by defining preconditional and postconditional states for state-chart graphs to capture semantics of concurrent objects to check commutativity. 
  
  \item We present a technique for representing concurrent programs in state-chart graphs using the graph database Neo4j. 
  
  \item We present ways to use the Cypher query language to issue queries over state-chart concurrent programs to study complex commutative structures and concurrency characteristics, like trace analysis, per thread execution analysis, and execution ordering of atomic actions. 
  
\end{itemize}

\section{Background}

In this section, we discuss in detail the rudimentary concepts that are required to understand our methodology in the representation of concurrent programs as graphs. We first discuss how transactions are associated with concurrent objects and the importance of identifying commutative operations to enhance performance in transactional execution. We then present trace theory for understanding preconditional and postconditional states in state-chart graphs. Finally, we present our tool, Concordant, which generates graphs from model checking data. 

\subsection{Asynchronous Transactional execution}

In terms of concurrent programming, a transaction is a composition of concurrent operations which appear to execute atomically and in isolation. Any transactional object is required to perform two types of operations to synchronize between each other: transactional synchronization and rollback/abort \cite{herlihy2008transactional} \cite{herlihy1993transactional}. Any two conflicting transactions that need to synchronize their executions require calling a mutual exclusion construct, for example a transactional lock. This strategy avoids collision by forecasting conflicting points. Rollback/abort is necessary when collisions or conflicts cannot be speculative, and upon collision a rollback operation is performed to undo the incomplete transactional changes. 

Two transactions, each with a set of methods that collectively commutes with one another, can execute concurrently without transaction-level synchronization. The commutative property is an important criteria to deduce speculative collision and improve performance in concurrent transactions. API designers can use Concordant to check which operations commute and under what circumstances while designing interfaces for transactional applications. Thus, the use of transactional locks can be greatly reduced by speculatively avoiding transactional synchronization using the commutative property at an object-level.

\subsection{Commutativity and Trace Monoids}

Although commutativity was formally presented in trace theory \cite{mazurkiewicz1995introduction}, its benefits in transactional boosting is a later discovery. We present a brief overview on the theory of traces that are used in formal analysis of concurrent computations using abstract symbols. This is useful to understand ways in which concurrent operations can commute with one another. Furthermore, we apply the same notion of commutativity into our example program to determine all possible commutative states. Commutative states are the basis for finding the preconditional and postconditional states in traces that helps in verifying if operations are commutable or not.   

A trace is a set of strings, where commutative letters represent the portion of a concurrent program that can execute independently, while non-commutative letters represent the portion of a concurrent program that cannot execute independently. With this idea, in trace theory all letters are broadly classified to be a \textit{dependency} (represented as \textit{D}) or an \textit{independency} (represented as \textit{I}). A dependency is any finite, symmetric and reflexive relation. Considering a finite order pair set \textit{(a,b)} in \textit{D}, then \textit{(a,a)} and \textit{(b,a)} are also in \textit{D} \cite{mazurkiewicz1995introduction}. A domain of D, represented as {$\sum_{D}$}, is defined as set of all alphabets used to define \textit{D}. For example, if \textit{D} is represented by all English lower case letters, then {$\sum_{D}$} is \{a,..,z\}. For simplicity, we confine our domain to only a few English lower-case characters. For a given dependency \textit{D}, an independency is the relation {$I_D$} = ({$\sum_{D}$} $\times$ {$\sum_{D}$}) - \textit{D}, which is symmetric and reflexive \cite{mazurkiewicz1995introduction}. All traces are primarily defined on the basis of \textit{dependencies}. 

A \textit{monoid} is an algebraic structure with an associative binary operation and an identity element \cite{mazurkiewicz1995introduction}. Let {$\sum_{}$} be an alphabet, then {$\sum_{}^*$} represents the set of all possible strings represented from {$\sum_{}$}. Here {$\sum_{}^*$} denotes the free monoid and * represents Kleene's operator. The independency relation yields an equivalence relation that partitions the free monoids into a set of equivalence classes, where the result is called a \textit{trace monoid} \cite{mazurkiewicz1995introduction}. The elements of the equivalence classes are the traces.

An independency relation \textit{I} yields a binary relation {$\sim$} on {$\sum_{}^*$}, where \textit{u} {$\sim$} \textit{v} if and only if \textit{x, y}{$\in$} {$\sum_{}^*$} and \textit{(a,b)} {$\in$} \textit{I} such that \textit{u = xaby} and \textit{v = xbay}. Here, \textit{x} and \textit{y} are preconditional and postconditional states, such that all commutative orderings of letters \textit{a} and \textit{b} happen in between \textit{x} and \textit{y}.  

Now let us consider the following example, let alphabet {$\sum_{}$} = \textit{\{a,b,c\}}. Assuming one possible dependency relation is 

\noindent\textit{D = {$\{a,b\}^2$} {$\cup$} {$\{a,c\}^2$}}

\textit{= \{a,b\} $\times$ \{a,b\} {$\cup$} \{a,c\} $\times$ \{a,c\}}

\textit{= \{(a,a), (a,b), (b,a), (b,b)\} {$\cup$} \{(a,a), (a,c), (c,a), (c,c)\}}

\textit{= \{(a,a), (a,b), (b,a), (b,b), (a,c), (c,a), (c,c)\}}

Then the corresponding independency is  {$I_D$}\textit{ = \{(b,c), (c,b)\}}. 
Clearly, letters \textit{b} and \textit{c} commute. Thus a trace over independency \textit{I} over string \textit{aabbca} is \big[\textit{aabbca}\big]{$_D$} = \{\textit{aabbca}, \textit{aabcba}, \textit{aacbba}\}. 

\begin{figure}
\centering
\includegraphics[width=0.35\textwidth]{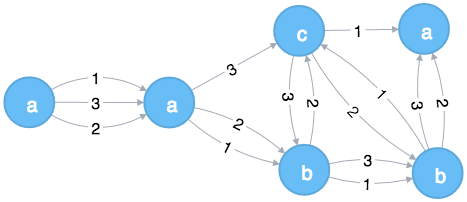}
\caption{State-chart representation for independency \textit{I}, over string \textit{aabbca} is \big[\textit{aabbca}\big]{$_D$} = \{\textit{aabbca}, \textit{aabcba}, \textit{aacbba}\}. }
\label{sample_full_graph}
\end{figure}

\begin{figure*}[htb]     
\centering
\begin{subfigure}{0.45\textwidth}\centering\includegraphics[width=0.74\columnwidth]{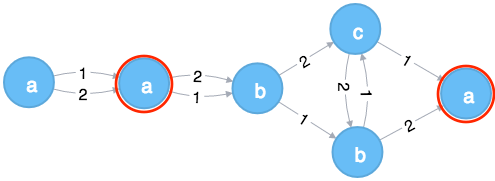}
\caption{State-chart graph for trace 1 and trace 2}
\label{sample_commutative12}
\end{subfigure}
\begin{subfigure}{0.45\textwidth}\centering\includegraphics[width=0.8\columnwidth]{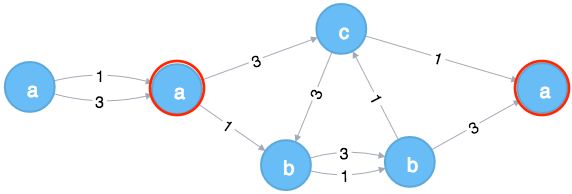}
\caption{State-chart graph for trace 1 and trace 3}
\label{sample_commutative13}\end{subfigure}\\
\begin{subfigure}{0.45\textwidth}\centering\includegraphics[width=0.8\columnwidth]{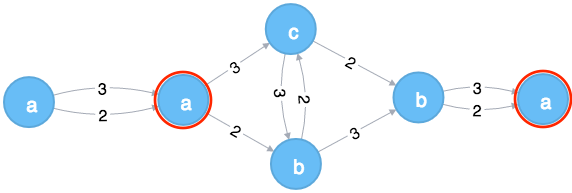}
\caption{State-chart graph for trace 2 and trace 3}
\label{sample_commutative23}
\end{subfigure}	
\caption{State-chart graph for analyzing commutativity induced by independency \textit{I} for trace \big[\textit{aabbca}\big]{$_D$}. Nodes highlighted with the red border represent preconditional and postconditional nodes. Preconditional and postconditional state nodes are defined in context with respect to how commutativity is observed in a graph. }
\label{sample_commutative_pair}
\end{figure*}

The notion of commutative operations through an independency relation is essential to understand the ways in which method invocations, responses, and atomic operations can be reordered with one another. Figure~\ref{sample_full_graph} depicts the graph representation for the above three traces, which shows all commutative relations between nodes for each trace pair. Nodes which are highlighted in Figure~\ref{sample_commutative_pair} represent the preconditional state followed by the postconditional state. Across all traces, these conditional state nodes remain constant, no matter how independent letters commute. In the following sections we check for similar preconditional and postconditional states in actual code execution to verify commutative operations. 

\section{Concordant}

Concordant is a graph generating tool that uses the Neo4j \cite{Neo4j} graph database platform. For our research, we use Neo4j for state-chart representation of concurrent program executions. A model checker is the crux of the Concordant tool.
To generate model checking data, Concordant uses the model checker CDSChecker \cite{norris2013cdschecker} to track all program states across multiple exhaustive thread interleavings. CDSChecker tracks model data properties like action types (atomic read/write), thread ID, memory orders (introduced in C11/C++11), memory addresses, and values associated with memory address, which are used to define states in our state chart. We customize CDSChecker to collect additional model checking information such as method invocation/response and method input/output. 

In this paper we reason about commutativity based on atomic action nodes in a state-chart, which is in itself a new approach. In the non-blocking programming paradigm, an atomic instruction ensures changes in shared resources/objects takes place in a race-free way. Thus, we assume atomic instructions are the basis of any modification in shared memory. Every recorded atomic instruction is depicted as an action node or state in our state-chart. In this context, a state contains information regarding an atomic instruction that is captured during model checking. Method invocations and responses are essential for determining all memory accesses that occur in a data structure, which is necessary to understand how two methods have ordered themselves in a trace. To check commutativity, we look for the same methods in different traces, reordered in different ways. We will discuss the actual execution runs in Section~\ref{Verification}, and inspect the model checking properties that are tracked using Concordant which describe the action nodes in the state-chart. Similar model checking instances across different traces are then merged into one single action state in the state-chart. Thus, this step reduces our search space generating a state-chart graph from independent trace results.

\section{Verifying Commutativity from State-Chart Notation of Concurrent Programs}
\label{Verification}

In this section, we dive into the concepts that are used to define action nodes and transitions in our state-chart graph, based on which we verify commutativity. We then present cypher queries to comprehend other characteristics of commutative operations from their graph notations.  

\subsection{States and Transitions}
\label{States}
In this section, we define our states or action nodes in state-charts. This corresponds to analyzing preconditions and postconditions between concurrent operations across different interleavings. As discussed earlier, we customize CDSChecker as our underlying model checker and we track the following attributes, to identify different states in our state chart: 
\begin{itemize}
\item \textbf{Action Type} - This entity stores information of different atomic instructions that takes place on shared objects, for instance, atomic read, atomic write, atomic read-modify-write(RMW), etc. We also use action types to store different stages of a thread's life-cycle in the program: thread create, thread start, thread join, and thread finish. Using custom annotations we also store method invocation and method response for analyzing real-time ordering of operations to compare object states. 

\item \textbf{Thread ID }- All action types are associated with a specific thread ID. For example, a specific thread will invoke a specific method or atomic operation. We link each state to a thread ID based on the thread invoking the action represented by the particular state. 

\item \textbf{Memory Location} - When verifying commutativity among method calls, it is important to know the active memory locations of a concurrent object in execution. An active memory location is a location in which an atomic instruction operates. 
The memory location attribute in action nodes helps to index the underlying object state before and after the invocation of methods. In order to reason about whether any ordering of two given operations results in the same effect in the underlying data structure, we compare the preconditional and postconditional action nodes of an individual trace. Any ordering between two operations that results in active memory location differing at object-level would lead to two distinct action nodes, from which we may infer that the operations do not commute. In corollary to that, operations that have the same active memory location at object-level would result in similar preconditional and postconditional action nodes when the operations successfully commute. 

\item \textbf{Method Input/Response Value} - Along with the memory location, the input/response value of atomic operations at that memory location helps to define a distinctive state in the state-chart. There could be some instance where scrutinizing only memory location might not be enough to infer commutativity across two or more operations. For instance, in a queue we are interested in checking the commutativity between enqueue(100) and enqueue(150). It is with the help of input values 0x64(which is hex for 100) and 0x96(which is hex for 150) that conveys which operations followed the other(enqueue(100) before enqueue(150), or vice-versa). Since reordering enqueue(100) and enqueue(150) will result in a different final state, we can reason from the state-chart that enqueue(100) and enqueue(150) do not commute. 

\item \textbf{Trace ID} - Every unique interleaving of a thread's course is a distinct trace which is specified by the trace ID. In our state-chart, trace ID is an attribute which we associate with transitions between action states. In our cypher query, we use conditions over this field to filter respective traces. 

\end{itemize}

We define preconditional and postconditional states (together \textit{conditional states}) based on action nodes in the state-chart graphs. Each action node is formed by merging trace actions across multiple exhaustive traces. These conditional states are associated with the state at which a data structure or object remains before and after method calls. Valid commutative methods reordered across different traces leave the same resultant effect on the underlying objects. Conditional states are dependency nodes, while action nodes generated during method reorderings are independencies. Thus, conditional states are the points at which commutativity is validated. In any trace analysis, the definition of preconditional and postconditional states is established based on the data structure and its operations that are considered for commutativity checks.

For example, in a set data structure, two add() operations can commute with each other if they are adding different elements to the set.
In this case, our preconditional state includes the action nodes ordered before both add() method invocations, and the postconditional state includes the action nodes ordered after the action nodes representing the response events of the two add() methods. 

\subsection{Commutativity in Concurrent Traces}

In this section we check for commutativity by analyzing the preconditional and postconditional states. Based on the type of concrete data structure used in designing the abstract objects, we analyze conditional states in two different approaches.

\subsubsection{Commutativity Verification on Non-Contiguous Memory: Linked List Based Set}
We use a set based on a concurrent linked list \cite{herlihy2011art} for reasoning about the commutative property in non-contiguous memory data structures. 
In non-contiguous memory allocation, each atomic operation takes place at different memory locations and thus, will generate different action nodes in the preconditional and postconditional states. We probe the underlying linked list with an add() operation such that the argument in the add() operation is greater than any existing value in the set. In doing this, the atomic read in the add() of the concurrent linked list checks for the value starting from the head node and stops at the tail node, where it finally performs an atomic write to insert the element into the set. If two operations commute with one another, the number of read action nodes in the postconditional state will remain the same across all traces.

\begin{figure*}
\centering
\includegraphics[width=0.85\textwidth]{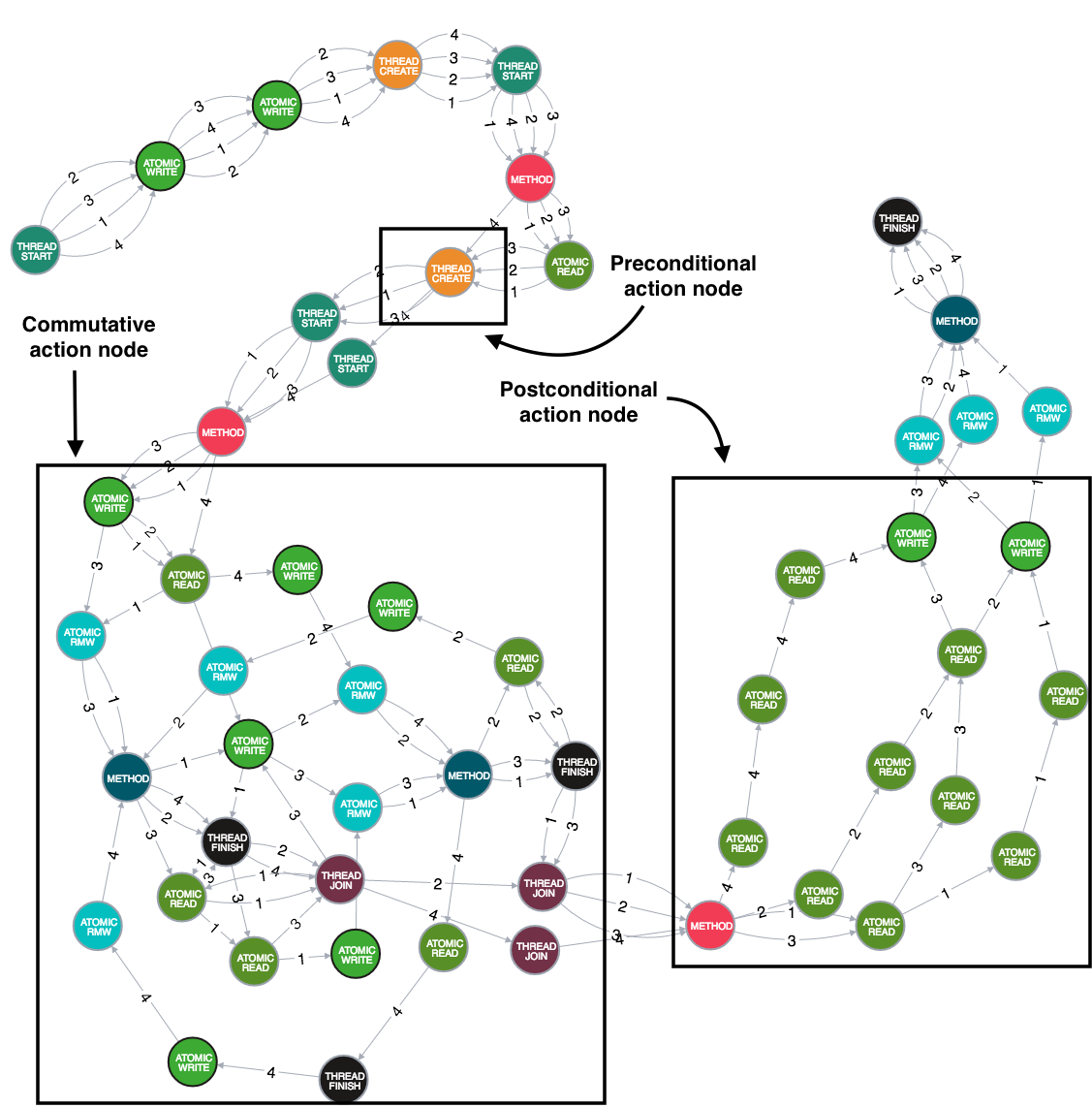}

\caption{\textbf{State-chart representation of a concurrent linked list\cite{herlihy2011art} based set showing two add() operations in four traces. Action nodes with similar action-types are represented in distinct colors. A method invocation is indicated by a pink action node. A method response is indicated by a blue action node. In non-contiguous memory, the number of read action nodes for each trace in the postconditional state is the same.}} 
\label{Commutativity_2_add_set}
\end{figure*}

Figure~\ref{Commutativity_2_add_set} shows an add($X$) operation and add($Y$) operation over a set object, where $X \ne Y$. The preconditional state is an empty set and the postconditional state includes an add($Z$) operation, where $Z > X$ and $Z > Y$. Since each trace yields the same number of read action nodes in the postconditional state, add($X$) and add($Y$) commute when $X \ne Y$.   

\begin{figure*}
\centering
\includegraphics[width=0.85\textwidth]{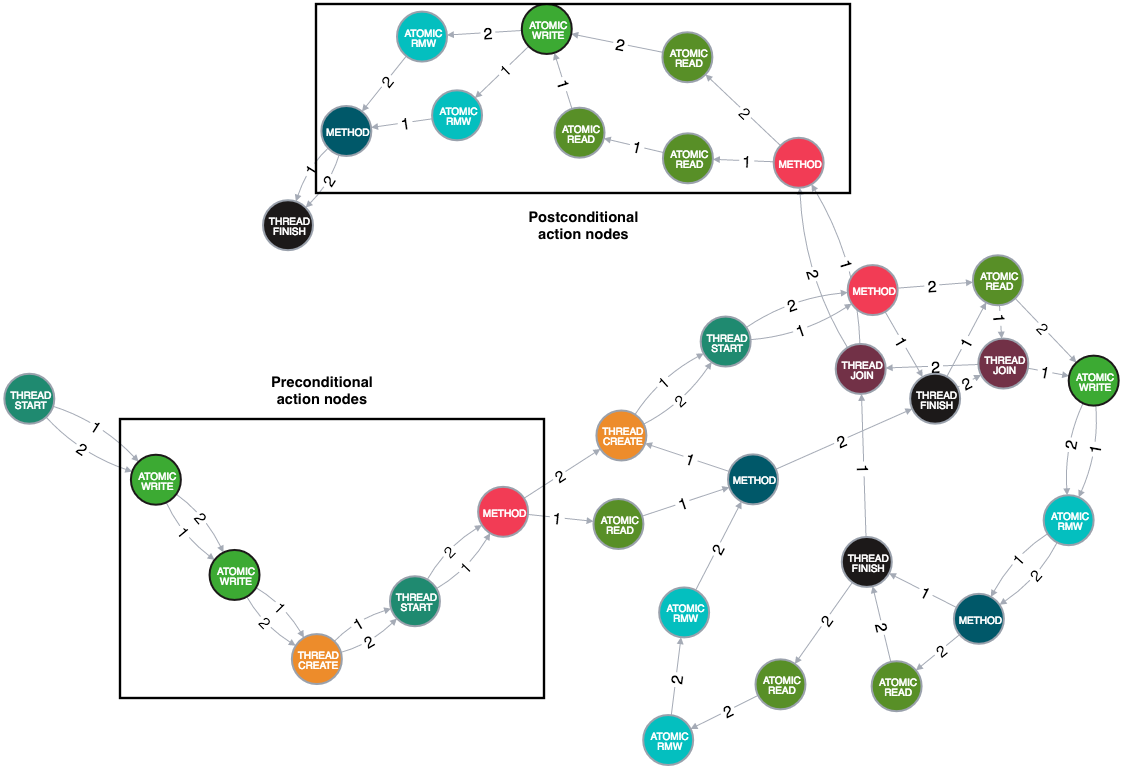}

\caption{\textbf{State-chart representation of a concurrent linked list\cite{herlihy2011art} based set showing one add() and one remove() operation in four traces. In non-contiguous memory, postconditional action nodes are not in concordance for all trace results.}} 
\label{nonCommutativity_set}
\end{figure*}

An example of two non-commutative methods is shown in Figure~\ref{nonCommutativity_set} for one add($X$) and one remove($X$) operation, where the precondition is an empty set. Similar to the previous example, the linked list is probed with an add($Z$) operation such that $Z > X$. The postconditional state demonstrates that trace 1 encounters one more atomic read than trace 2. Trace 1 encompasses the history in which remove($X$) is ordered before add($X$). Since the list is initially empty, the remove($X$) will fail and add($X$) will successfully add $X$ to the list, resulting in a list with one element in the postconditional state. Trace 2 encompasses the history in which add($X$) is ordered before remove($X$). Add($X$) will successfully add $X$ since the list is initially empty, and remove($X$) will successfully remove $X$, resulting in an empty list in the postconditional state. Since the atomic operations invoked in the postconditional state are not identical for the two traces, we verify that add($X$) and remove($X$) are not commutative. 

\subsubsection{Commutativity Verification on Contiguous Memory: Array-Based Queue}

In an array-based queue, the array is the concrete data structure that allocates memory in contiguous locations during program run-time. This restrictive memory utilization induces action nodes to have identical memory maps across different traces. 

\begin{figure*}
\centering
\includegraphics[width=0.65\textwidth]{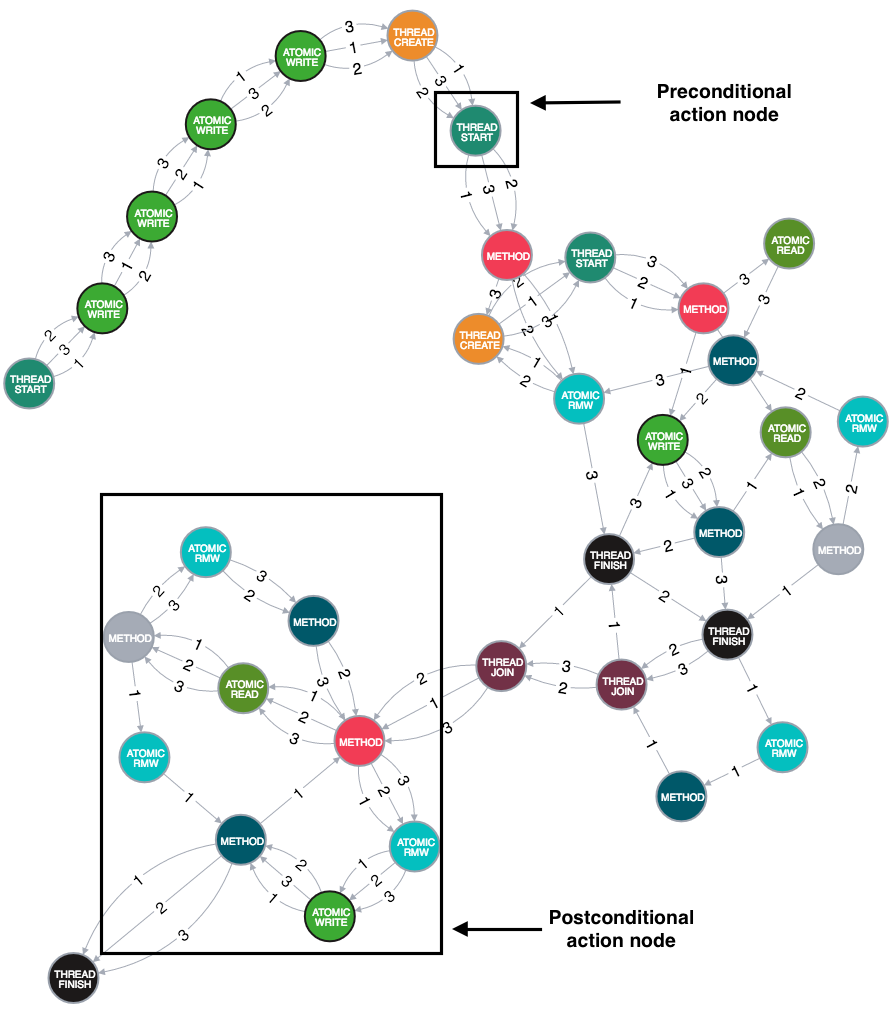}

\caption{State-chart representation of Herlihy-Wing Queue \cite{herlihy1990linearizability} showing one dequeue() and one enqueue() operation in three traces. Since the queue is empty in the precondition, we consider the first THREAD START as the preconditional action node. Postconditional action nodes are not consistent across all traces. } 
\label{non_Commutativity_2_deq_HWQ}
\end{figure*}

Our commutativity analysis for contiguous memory is on the Herlihy-Wing Queue \cite{herlihy1990linearizability}. 
Queues differ from sets because the order in which elements are enqueued affects the final object state and the order in which elements are dequeued affects the response of the operations. The enqueue() and dequeue() operations are commutative with each other as long as the queue is not empty. 
The output model checking data for a scenario in which an enqueue() and dequeue() operation are invoked on an empty queue is shown in Figure~\ref{non_Commutativity_2_deq_HWQ}. The preconditional state on the object is an empty queue. 
We probe the postconditional state of the queue structure with a dequeue() followed by an enqueue() operation executed sequentially. 
If the enqueue() is ordered before the dequeue() in the concurrent execution, the queue will be empty and the probed dequeue() will not perform an atomic read, as shown in the traces not containing the atomic read in the postcondition state. If the dequeue() is ordered before the enqueue() in the concurrent execution, the queue will contain one element and the probed() dequeue() will perform an atomic read on the element in the queue, as shown in the traces containing the atomic read in the postconditional state. 
Clearly, the action states in the postconditional states are not consistent in all the traces, indicating the enqueue() does not commute with dequeue() when the queue is empty.  

\subsection{Beyond Commutativity: Concurrent Trace Analysis using Cypher Query Language}

In this section, we discuss identifying commuting actions to map them to the instructions of their respective methods. The advantage of using graphs in analyzing concurrent programs is that in addition to checking whether methods commute, we can also analyze how they commute during model checking. For our concurrent trace analysis over state-chart graphs we use graph database tool Neo4j and the Cypher query language. 
Based on the model checking attributes mentioned in Section~\ref{States}, we define properties of a graph node in the graph database. We also make use of the trace ID to uniquely identify the properties of relationships between nodes. In our graph database, nodes represent states in the state-chart, and the relationships represent the directed transitions in the state-chart. 

\begin{figure*}
\includegraphics[width=\textwidth]{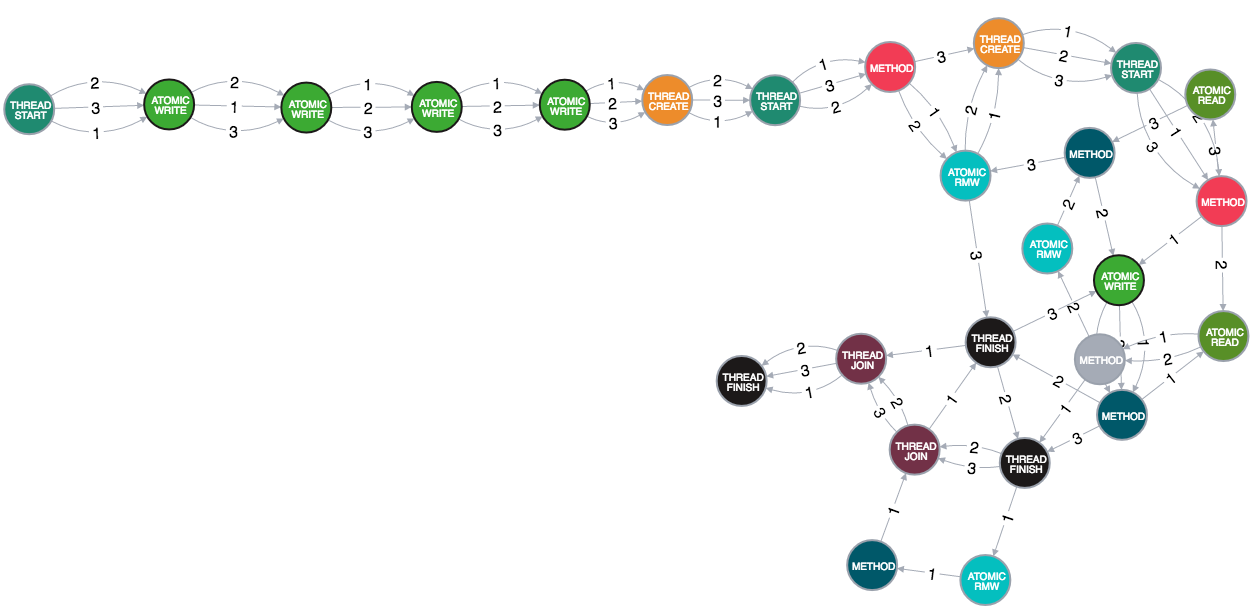}
\caption{State-chart representation of Herlihy-Wing Queue \cite{herlihy1990linearizability} showing one enqueue() and one dequeue() operation in two separate threads. Action nodes with similar action-types are represented in distinct colors. Each transition is labeled with a trace ID to keep track of their execution sequence in the program. All state-charts start with the \textit{THREAD START} action node as the initial state, which is the start of main thread in the actual program. Every complete graph that runs without an error ends at the action node \textit{THREAD FINISH}. }
\label{HWQ_full_graph}
\end{figure*}

\begin{figure*}
\includegraphics[width=\textwidth]{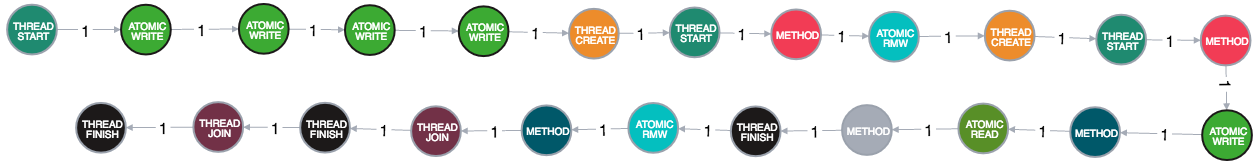}
\caption{Trace 1 of state-chart representation of one enqueue() and one dequeue() in the Herlihy-Wing Queue.}
\label{HWQ_trace1_graph}
\end{figure*}

\begin{figure*}
\includegraphics[width=\textwidth]{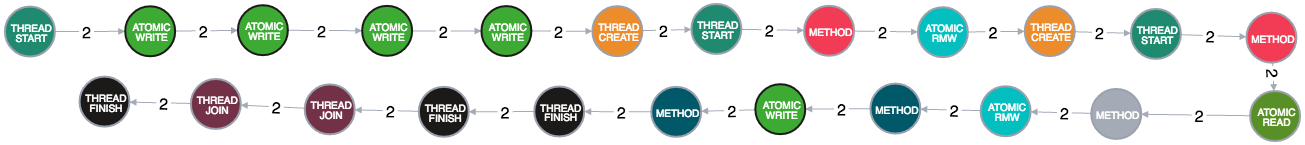}
\caption{Trace 2 of state-chart representation of one enqueue() and one dequeue() in the Herlihy-Wing Queue. }
\label{HWQ_trace2_graph}
\end{figure*}

\begin{figure*}
\includegraphics[width=\textwidth]{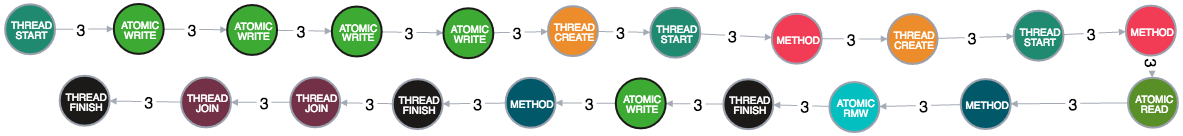}
\caption{Trace 3 of state-chart representation of one enqueue() and one dequeue() in the Herlihy-Wing Queue.}
\label{HWQ_trace3_graph}
\end{figure*}

Figure~\ref{HWQ_full_graph} is a sample state-chart graph of the Herlihy-Wing Queue, with one enqueue() and one dequeue() operation. Each of these operations are forked from two separate threads. In addition, there is a main thread that initializes atomic variables followed by spawning threads. We use Concordant to generate Cypher queries to create action nodes and relationships for generating the corresponding graph. In this particular example there are three different ways these two queue operations can be reordered. 
Figures \ref{HWQ_trace1_graph}, \ref{HWQ_trace2_graph}, and \ref{HWQ_trace3_graph} show the history of each interleaving. 
The variable $r$ denotes a relationship in the Cypher query language. The variables $a$, $b$, and $c$ denote nodes in the graph database. We use the following query to generate the respective graphs: 

\begin{lstlisting}[language=SQL]

MATCH (a)-[r]->(b) WHERE 1 IN r.id 
		      RETURN a,r,b
MATCH (a)-[r]->(b) WHERE 2 IN r.id 
                      RETURN a,r,b
MATCH (a)-[r]->(b) WHERE 3 IN r.id 
		      RETURN a,r,b
\end{lstlisting}

The \textit{id} property in the relationship between action nodes is used to store the respective trace information. Each action node contains thread properties that help filter operations executed thread-wise across a trace. In a complete execution history of a concurrent program, the ordering of operations carried out by multiple threads vary based on ways they could interleave. For example, in a typical history of a concurrent queue execution, the invocation of an enqueue() operation by thread 1 may be followed by the invocation and response of a dequeue() by thread 2, closing with a response of enqueue() by thread 1. In this example trace, thread 2 has two consecutive action states whereas thread 1 has two nonconsecutive action states. It is important to evaluate traces from this perspective to formulate a cypher query to fetch all specific action nodes. 

In the following cypher, we fetch the thread specific operations using these two cases: The first case (a) - [r] -\textgreater (b) fetches all action nodes and their relationships that are executed by one thread (WHERE a.Thread=`1' AND b.Thread=`1') in a particular trace (say 1 IN r.id), where more than one operation occurs without switching to another thread. The second case fetches all singly occurred action nodes (c) which were not fetched in the first case, WHERE c.Thread=`1' of the same trace. We perform this cypher query for all three threads.  

\begin{lstlisting}[language=SQL]
MATCH (a)-[r]->(b),(c) 
WHERE 1 IN r.id AND a.Thread='1'
AND b.Thread='1' AND c<>a AND c<>b 
AND c.Thread='1' RETURN a,r,b,c

MATCH (a)-[r]->(b),(c) 
WHERE 1 IN r.id AND a.Thread='2'
AND b.Thread='2' AND c<>a AND c<>b 
AND c.Thread='2' RETURN a,r,b,c

MATCH (a)-[r]->(b),(c) 
WHERE 1 IN r.id AND a.Thread='3'
AND b.Thread='3' AND c<>a AND c<>b 
AND c.Thread='3' RETURN a,r,b,c
\end{lstlisting}

We can do a similar query for trace 2 by replacing `WHERE 1 in r.id' with `WHERE 2 in r.id' in the previous query. Similarly, the same query can be performed for trace 3 by replacing `WHERE 1 in r.id' with `WHERE 3 in r.id' in the previous query.   

\begin{figure*}[htb]     
\centering
\begin{subfigure}{\textwidth}\centering\includegraphics[width=\columnwidth]{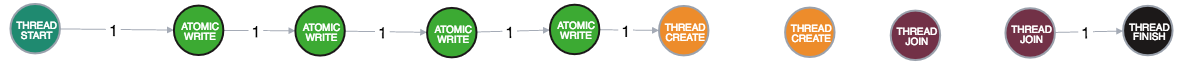}\caption{List of operations executed by Thread 1 in Trace 1}
\label{HWQ_trace1thread1_graph}
\end{subfigure}
\begin{subfigure}{0.8\textwidth}\centering\includegraphics[width=.7\columnwidth]{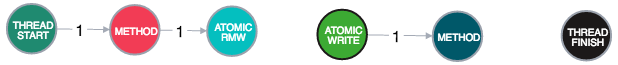}\caption{List of operations executed by Thread 2 in Trace 1}
\label{HWQ_trace1thread2_graph}
\end{subfigure}\\
\begin{subfigure}{0.8\textwidth}\centering\includegraphics[width=0.75\columnwidth]{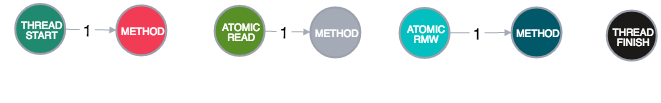}\caption{List of operations executed by Thread 3 in Trace 1}
\label{HWQ_trace1thread3_graph}
\end{subfigure}
\caption{Trace 1 - Per thread analysis of one enqueue() and one dequeue() in the Herlihy-Wing Queue}
\label{HWQ_trace1perthread_graph}
\vspace{-1em}
\end{figure*}

In order to identify the commutative states we need to map each instruction to their respective methods. 
For trace analysis, we use the following points to facilitate our reasoning on identifying action node sets that can be used to check commutative methods. 
\begin{itemize}

\item We use the main thread (thread 1) to analyze if an executing graph has terminated with or without any error. Its purpose is to fork out threads that carry out actual operations. Therefore, the main thread is used for establishing preconditional and postconditional states and can be ignored in checking commutative action nodes.

\item Each method call is composed of one or more atomic instructions. When we query on a method called by a thread specific to a trace, we see the same set of atomic action nodes in the resultant graph as that in the source code method call.

\end{itemize}

Figures \ref{HWQ_trace1_graph}, \ref{HWQ_trace2_graph}, and \ref{HWQ_trace3_graph} are the state-chart graphs for traces 1, 2, and 3, respectively. From Figure~\ref{HWQ_trace1_graph} we see there are five atomic write operations, two atomic RMW operations and one atomic read operation. Referring to Figures \ref{HWQ_trace1_graph}, \ref{HWQ_trace2_graph}, and \ref{HWQ_trace3_graph}, we see that four atomic writes executed by thread 1 are common across all traces. Referring to our implementation of the Herlihy-Wing Queue, we see that these write operations refer to the initialization of the atomic variables carried out by the main thread at the start of the program. Since these steps take effect sequentially by the main thread, in Figure~\ref{HWQ_trace1thread1_graph} we see a consistent interleaving of action nodes across all the traces. Referring to the source code, we see that the main thread (thread 1), spawns out the other threads by calling the thread create instruction and remains inactive until the other threads finish their tasks, and resumes back on the thread join instruction followed by finishing itself at thread finish. Referring to thread 2's trace in Figure~\ref{HWQ_trace1thread2_graph}, we see action nodes atomic RMW and atomic write instructions, which represent the enqueue(). Referring to thread 3's trace in Figure~\ref{HWQ_trace1thread3_graph}, we see action nodes atomic read and atomic RMW instructions, which represent the dequeue(). 

\section{Related Work}

Our research on commutativity using graphs was inspired by the work done by Antoni Mazurkiewicz in trace theory\cite{mazurkiewicz1995introduction}. Mazurkiewicz presented concurrent processes as abstract strings that are also used in representing sequential systems. By representing concurrent strings as dependency and commuting independency characters, he devised dependency graphs to derive traces, that sowed our foundation for graphical representation of concurrent traces. 

Using commutativity as a means to increase transactional concurrency has been a popular technique~\cite{hassan2014optimistic,spiegelman2016transactional,zhang2018lock,laborde2019wait}. Transactional boosting~\cite{herlihy2008transactional} enables commutative operations in a transaction to proceed concurrently and provides transactional synchronization only for non-commutative operations through fine-grained abstract locks.
Lock-Free Transactional Transformation~\cite{zhang2016lock} proposes a lock-free transactional synchronization approach for non-commutative operations while eliminating the need for a physical rollback through a logical interpretation of the correct abstract state.
Commutativity has also been exploited in verification tools~\cite{shacham2011testing,peterson2017transactional} by pruning a reordering of commutative operations from the search space when generating the sequential specification for a concurrent execution.

Current understanding of transactional boosting has evolved though numerous investigations and diversified research done towards understanding object-level commutativity. Steele et al.\cite{steele1989making} described commutativity as conflict-free operations. Thus, operations that could be analyzed for conflict-freedom can be inferred as commutative to each other. Prabhu\cite{prabhu2011commutative} and Rinard\cite{rinard1997commutativity} describe ways to synchronize operations to ensure that changes occur  compositely, even when memory locations are conflicting. They have emphasized on the correctness over scalability. In contrast to this approach, our focus is only on scalability, as we have relied on the underlying concrete data structures for program correctness. Analysis of commutativity based on memory state during run-time has been used to comprehensively describe commutative operations \cite{aleen2009commutativity}. In our case, we use a model checker to keep track of memory locations and the response values of method calls. Object-based commutative analysis depending on observing method reordering patterns has been a popular approach \cite{rinard1997commutativity}\cite{eberhard2005object}. Previous to their work, commutativity was observed on read/write operations at memory word level. We extend object-based commutative analysis to generate graphs and interconnections between trace flows. 

Clements et al. \cite{clements2015scalable} present Commuter, a conflict checking tool that lists out different combinations of operations that can commute at run-time. Commuter and our tool are both based on conditional states, but Commuter does not track object-level model checking data. Such data is necessary to verify that the abstract state of the data structure remains the same in checking commutative operations. Clements' tool is an inspiration in the development of our methodology to check commutativity in object-level operations. 

\section{Conclusion}

We have presented Concordant, a graphical representational tool to generate graphs to check commutative operations during model checking. By representing model checking data in state-chart action nodes, we are able to find preconditional and postconditional states as a base to reason about commutativity. Our work is the first application of graph database in analyzing concurrent programs by using Cypher query language. This is not only an intuitive approach to reason about commutativity between operating methods but also an extensible approach to check other properties of concurrent programs.

\bibliographystyle{ACM-Reference-Format}
\bibliography{sigproc} 

\end{document}